\documentclass[journal,twoside,web]{ieeecolor}
\usepackage{tmi}
\usepackage{cite}
\usepackage{amsmath,amssymb,amsfonts}
\usepackage{algorithmic}
\usepackage{graphicx}
\usepackage{textcomp}

\usepackage{booktabs}
\usepackage{caption}
\usepackage{url}
\usepackage{pifont}
\usepackage{multirow}
\usepackage{subfig}
\usepackage{xurl}

\def\BibTeX{{\rm B\kern-.05em{\sc i\kern-.025em b}\kern-.08em
    T\kern-.1667em\lower.7ex\hbox{E}\kern-.125emX}}
\begin{document}
\title{\emph{RankByGene}: Gene-Guided Histopathology Representation Learning Through Cross-Modal Ranking Consistency}
\author{Wentao Huang, Meilong Xu, Xiaoling Hu, Shahira Abousamra, Aniruddha Ganguly, Saarthak Kapse, Alisa Yurovsky, Prateek Prasanna, Tahsin Kurc, Joel Saltz, Michael L. Miller, Chao Chen
\thanks{}
\thanks{Wentao Huang, Meilong Xu, Aniruddha Ganguly are with the Computer Science Department, Stony Brook University, Stony Brook, NY 11790 USA (e-mail: wenthuang@cs.stonybrook.edu, meixu@cs.stonybrook.edu, aniganguly@cs.stonybrook.edu).}
\thanks{Xiaoling Hu is with the Athinoula A. Martinos Center for Biomedical Imaging,
Massachusetts General Hospital and Harvard Medical School, Boston, MA 02115 USA (e-mail: xihu3@mgh.harvard.edu).}
\thanks{Shahira Abousamra is with the Department of Biomedical Data Science,
Stanford University, Stanford, CA 94305 USA(e-mail: shsamra@stanford.edu).}
\thanks{Saarthak Kapse, Alisa Yurovsky, Prateek Prasanna, Tashin Kurc, Joel Saltz, Chao Chen are with the Department of Biomedical Informatics, Stony Brook, NY 11790 USA (e-mail: saarthak.kapse@stonybrook.edu, alisa.yurovsky@stonybrook.edu, prateek.prasanna@stonybrook.edu, tahsin.kurc@stonybrook.edu, Joel.Saltz@stonybrookmedicine.edu, chao.chen.1@stonybrook.edu).}
\thanks{Michael L. Miller is with the Department of Pathology and Cell Biology, Columbia University, New York, NY 10027 USA (e-mail: mlm2326@cumc.columbia.edu).}
}

\newcommand{\myparagraph}[1]{\smallskip\noindent\textbf{#1}}

\maketitle

\begin{abstract}
Spatial transcriptomics (ST) provides essential spatial context by mapping gene expression within tissue, enabling detailed study of cellular heterogeneity and tissue organization. However, aligning ST data with histology images poses challenges due to inherent spatial distortions and modality-specific variations. Existing methods largely rely on direct alignment, which often fails to capture complex cross-modal relationships. To address these limitations, we propose a novel framework that aligns gene and image features using a ranking-based alignment loss, preserving relative similarity across modalities and enabling robust multi-scale alignment. To further enhance the alignment's stability, we employ self-supervised knowledge distillation with a teacher-student network architecture, which serves as an intra-modal stability regularizer that prevents image-representation drift during cross-modal alignment.  Extensive experiments on seven public datasets that encompass gene expression prediction, slide-level classification, and survival analysis demonstrate the efficacy of our method, showing improved alignment and predictive performance over existing methods. Code is available at \url{https://github.com/winston52/RankByGene}.
\end{abstract}


\section{Introduction}
\label{sec:introduction}

\IEEEPARstart{D}{igital} pathology has advanced significantly in recent years, thanks to the availability of large number of digitized slides and rapid development of deep learning. Powerful learning methods have been proposed for image-based prediction at different scales, including individual cells \cite{graham2019hover, horst2024cellvit}, larger structures like glands \cite{isensee2021nnu, wang2022uctransnet}, regions of interest \cite{le2020utilizing, ou2022patcher, deng2024prpseg, dominguez2024systematic}, and the whole slide \cite{ilse2018attention,li2021dual,lu2021data,kapse2024si,chen2021multimodal}. However, despite their promise, these methods are inherently constrained by the information that can be extracted from a digital slide. Bridging the gap between cell morphology, contextual information, and actual functionality remains a critical challenge, one that could be key to advancing image-based diagnosis and prognosis.

\begin{figure*}[t]
\centering
\includegraphics[width=0.9\textwidth]{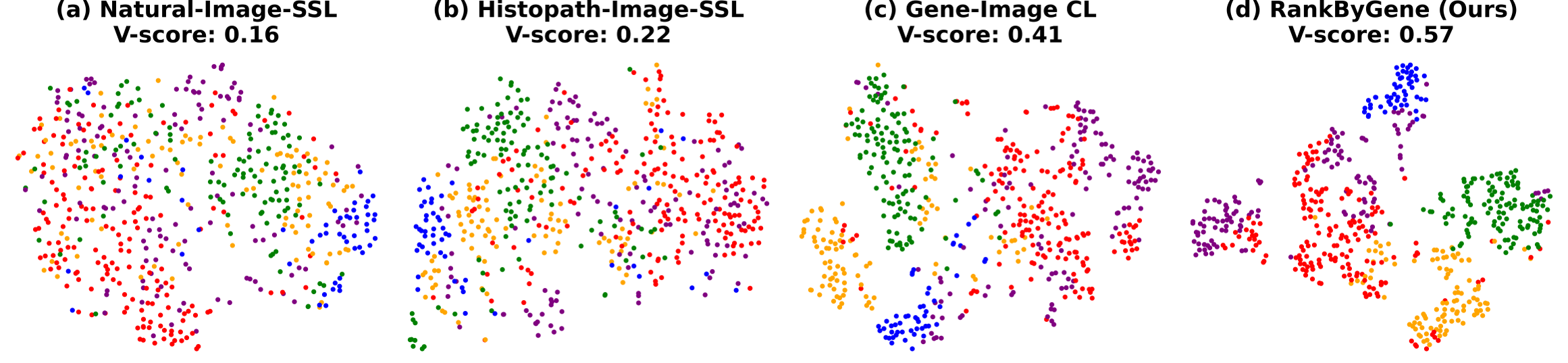}
 \vspace{-.1in}
\caption{
t-SNE \cite{van2008visualizing} visualization of image features of different spots in an ST slide. 
We show the feature learned using different methods, including (a) SSL on natural images, (b) SSL on histopathology images, (c) CL on ST data using InfoNCE loss \cite{jaume2024hest}, and (d) \emph{RankByGene}. 
Learning with gene information (c) clearly outperforms learning with images alone ((a) and (b)). Furthermore, our method (d) achieves even greater improvement, demonstrating better inter-class separability than (c), thanks to the proposed contributions.
Each spot is assigned a label via K-means clustering on gene expression values; the corresponding image features are then color-coded by these labels. 
We also provide quantitative measures using v-score \cite{rosenberg2007v}, where a higher value indicates better alignment of image features with gene expression.
}
 \vspace{-.1in}
\label{fig:tsne}
\end{figure*}

Transcriptomics \cite{wang2009rna} quantifies gene expression across tissues, providing a snapshot of cellular heterogeneity. Single-cell RNA sequencing \cite{tang2009mrna} refines this by profiling individual cells, enabling precise cell characterization but disrupting tissue architecture and losing spatial context \cite{saliba2014single}.
Spatial transcriptomics (ST) overcomes this limitation by preserving spatial organization while mapping gene expression \cite{rao2021exploring}. It captures cellular states and molecular interactions within their native environment \cite{marx2021method}, crucial for understanding disease dynamics and therapeutic responses \cite{moffitt2018molecular, ding2022image}. For instance, ST reveals distinct molecular signatures in tumor regions, such as the core, invasive margins, and immune-infiltrated zones, offering deeper insights into tumor heterogeneity \cite{ji2020multimodal}. However, its high cost remains a significant barrier to widespread clinical adoption.

Despite the high cost, ST data provides new opportunities to enhance existing affordable image-based prediction methods with novel biological information. Most ST data have both localized gene expressions and their corresponding histology image patches. One can exploit the two complementary modalities at a fine scale, unveiling the subtle visual cues from cell morphology and context that capture cell functionality.
As shown in recent studies \cite{chen2024stimage, jaume2024hest, min2024multimodal, xie2024spatially}, visual information including the density and spatial organization of various cell types (e.g., epithelial, lymphocyte, fibroblast) correlates with gene pathways and expressions. If correctly distilled, these gene-correlated visual cues can be valuable for downstream tasks.

In this paper, we study the problem of \emph{learning gene-guided image representations through ST data}. In particular, we use the gene expression to better align image features, so that these image features encode cell-functionality-specific visual information. 
As we will demonstrate in experiments, these gene-guided image features will improve image-based prediction of clinical outcomes. The learned features can be applied broadly to any image-only cohort, benefiting the digital pathology community.

Despite its scientific potential, multi-modal representation learning with ST data is challenging for various reasons. First and most importantly, it is fundamentally unknown how well the image and gene expressions could align. Different modalities may contain different information; ensuring complete alignment can be difficult or even impossible \cite{huh2024position}. Secondly, the image and gene features are encoded with different networks and have different initial representation powers; while images can be encoded with CNNs and can be initialized with powerful foundational models \cite{wang2021transpath,chen2022scaling,chen2024uni,xu2024whole,wang2024pathology,vorontsov2024foundation}, typically gene feature encoding can only be based on fully-connected networks and trained from scratch \cite{xie2024spatially,jaume2024hest,min2024multimodal}.
Furthermore, the gene expression data is extremely high-dimensional (with more than 15,000 genes per sample) and very sparse (on average, 80\% of gene expression values are zeros \cite{jaume2024hest}). Finally, as a novel technology, available ST data often suffers from experimental artifacts, such as spatial dependent noise \cite{abrar2023novatest} and sparsity \cite{li2022sd2,li2023comprehensive}.

To address these challenges, methods have been proposed to map both image patches and gene expressions into a shared latent feature space. Self-supervised contrastive learning losses have been adopted to ensure the image features and gene features from the same location (called a \emph{spot}) are mapped together~\cite{jaume2024hest}. The standard contrastive loss, however, overlooks the potential similarity between different spots. Xie et al.~\cite{xie2024spatially} employ a smoothed variant of the CLIP loss~\cite{radford2021learning} to encourage spots with similar image patches or similar gene expressions to be mapped closer in the feature space. Similarly, Min et al.~\cite{min2024multimodal} adopt a similar approach to enhance feature alignment. These methods, however, do not fully address the aforementioned challenges and lack an effective global strategy to align multi-modal features.

\begin{figure*}[t]
\centering
\includegraphics[width=.9\textwidth]{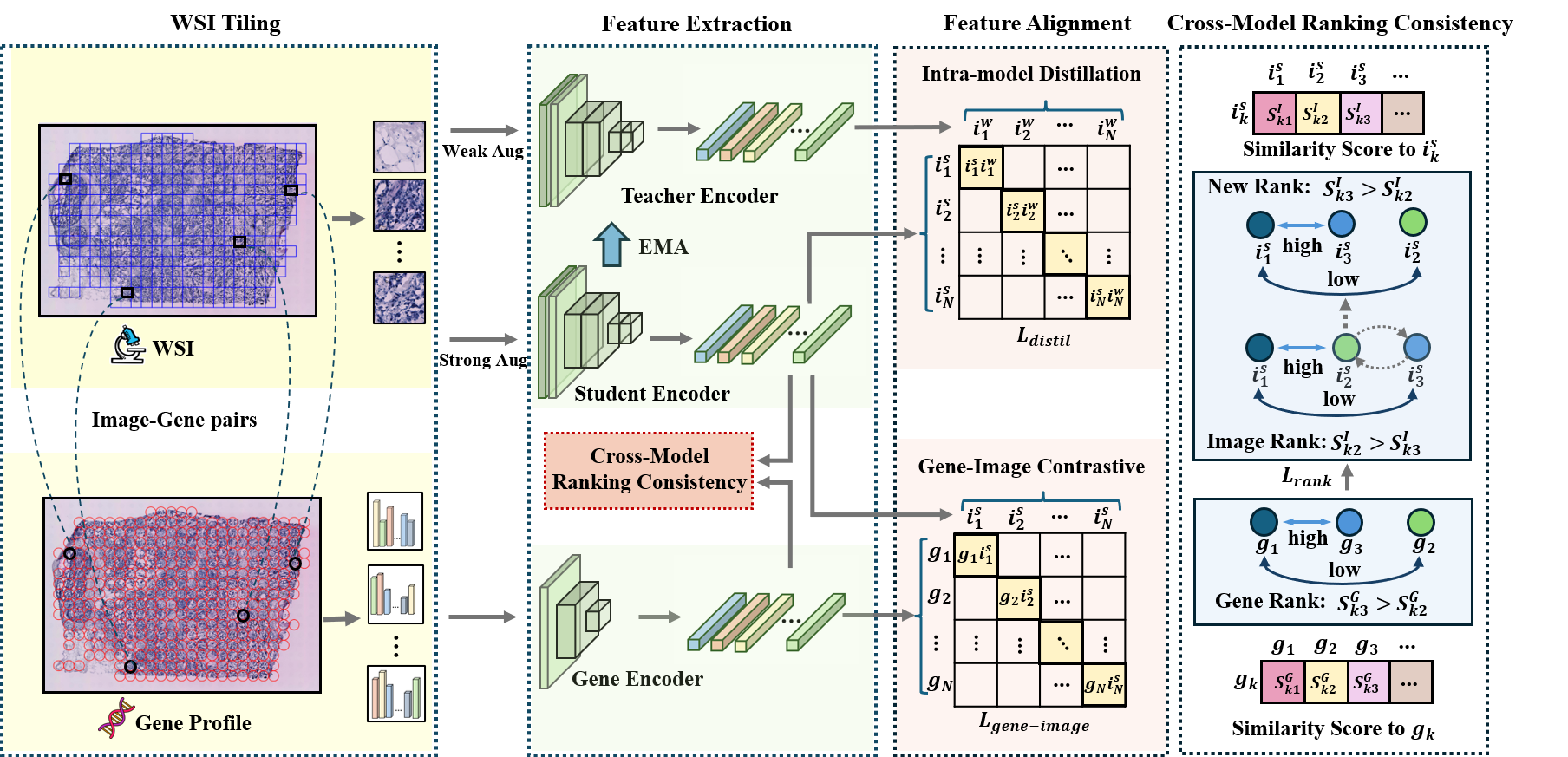}
 \vspace{-.05in}
\caption{Overview of our \emph{RankbyGene} framework. The framework begins with WSI Tiling, where WSIs are cut into patches, each paired with a gene spot. In feature extraction, weak and strong augmentations of the patches are processed through a teacher and student encoder, while a gene encoder extracts features from the gene profile. The feature alignment stage ensures that weakly and strongly augmented image features are aligned through intra-modal distillation loss and the image and gene features are aligned using gene-image contrastive loss. Meanwhile, our proposed cross-modal ranking consistency loss maintains consistent similarity ranking across two modalities. }
 \vspace{-.1in}
\label{fig:overview}
\end{figure*}

\myparagraph{\emph{RankByGene}.}
We propose \emph{RankByGene}, a novel approach to align image and gene features across multiple scales, both locally and globally. The alignment method is designed to be robust to modality-specific distortions, addressing the challenges outlined earlier. Specifically, we introduce an innovative ranking-based alignment loss ensuring that similarity relationships between gene expression features are reflected in the corresponding image features. 
 
This ranking-based approach offers several key advantages. First, it facilitates alignment across both local and global scales, enabling even distant features to interact and align. This global alignment complements the existing local matching of image and gene features of the same spot~\cite{xie2024spatially, min2024multimodal}. Second, by focusing on matching rankings rather than exact similarity values, our method is more robust to distortions, allowing it to tolerate similarity variations, especially between features that are far apart from one another.

To further enhance the robustness of the alignment, we adopt a self-supervised knowledge distillation approach to stabilize the learning of gene-informed image representations. By utilizing a teacher-student network architecture with both weakly and strongly augmented image patches, we ensure that the student network effectively learns the aligned features. This intra-modal regularizer encourages the image-branch representation to remain stable under augmentation and EMA teacher-student updates, providing a robust anchor for the cross-modal ranking signal. 

The strength of our method is demonstrated in Figure \ref{fig:tsne}, which shows the t-SNE visualization \cite{van2008visualizing} of image features from various methods: natural image-only self-supervised learning (SSL), histopathology image-only SSL, gene-image contrastive learning (CL) using InfoNCE \cite{jaume2024hest}, and \emph{RankByGene}. Gene-image CL clearly obtains image features better aligned with gene expressions. Our approach further boosts the features using the proposed contributions. 

In summary, we propose a novel multi-modal feature alignment method that integrates pathology images and ST data, enabling robust gene-informed image representation learning. Our contributions are threefold:
\begin{itemize}
\item We propose a novel cross-modal ranking consistency mechanism that enhances gene-image alignment at both local and global scales, while maintaining robustness to long-range distortions.

\item We introduce a knowledge distillation module to the gene-image alignment framework, which acts as an intra-modal stability regularizer on the image branch and prevents representation drift during cross-modal alignment optimization.

\item Extensive experiments on gene expression prediction, gene-related slide-level classification, and survival analysis tasks show that our method outperforms existing ones, underscoring its robustness and efficacy.
\end{itemize}

\section{Related Work}

\subsection{Histopathology Image Analysis}
Histopathology image analysis is regarded as the gold standard for cancer diagnosis and treatment \cite{niazi2019digital,barisoni2020digital}. Due to the huge resolution of whole slide images (WSIs) and the difficulty of obtaining patch-level labels, WSI analysis is commonly performed in a weakly supervised setting \cite{dietterich1997solving}, where only slide-level labels are available. Typical WSI analysis tasks include classification \cite{ilse2018attention,li2021dual,zhang2022dtfd,shao2021transmil,qu2023boosting,tang2023multiple,zhang2023attention}, which predicts the presence or absence of a tumor or identifies specific tumor subtypes, and survival prediction \cite{chen2021multimodal,xu2023multimodal,zhou2023cross,zhang2024prototypical,song2024multimodal, xu2025distilled}, which estimates patient mortality risk. Multiple instance learning (MIL) \cite{ilse2018attention,li2021dual,qu2022bi,lin2023interventional,zhu2025dgr,zhou2024pathm3} based frameworks have been proposed to model relationships between patches more effectively, enabling more accurate slide-level predictions.

\begin{figure*}[t]
\centering
\includegraphics[width=.75\textwidth]{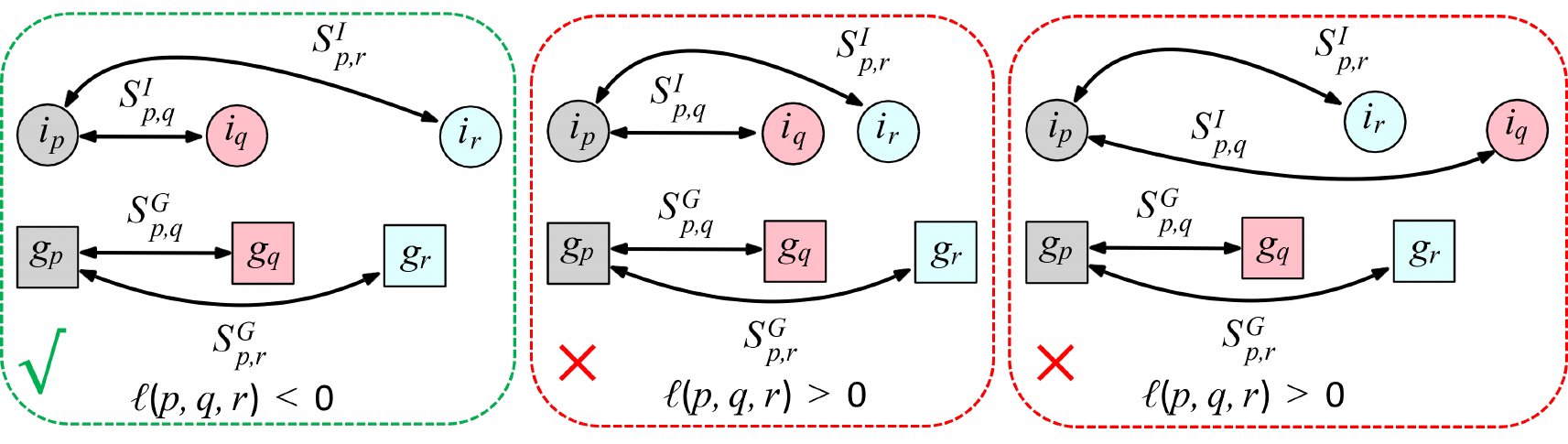}
 \vspace{-.1in}
\caption{Illustration of the ranking loss intuition, when the gene features $g_q$ are closer to $g_p$ than $g_r$. Note the similarity is inversely proportional to the distance. \textbf{Left:} $i_q$ is also closer to $i_p$ than $i_r$, and furthermore, and the gap between image feature similarities $S_{p,q}^I$ and $S_{p,r}^I$ is bigger than the gap between gene feature similarities $S_{p,q}^I$ and $S_{p,r}^I$. This is the desirable case where $\ell(p,q,r)$ is negative. \textbf{Middle:} when the similarity ranking is the same between gene and image features, but when the gap between image feature similarities is smaller than the gap between gene feature similarities. $\ell(p,q,r)$ is positive, and will incur penalty. \textbf{Right:} when the similarity order is inconsistent, $\ell(p,q,r)$ is positive. Undesirable.}
 \vspace{-.1in}
\label{fig:ranking}
\end{figure*}

\subsection{Image Feature Learning in Histopathology}
The performance of WSI analysis heavily relies on image feature learning. With advancements in self-supervised learning (SSL) in computer vision, methods like SimCLR \cite{chen2020simple}, MoCo \cite{he2020momentum}, and DINO \cite{caron2021emerging} have been adapted for image feature learning in histopathology, leading to the development of various foundation models pretrained on WSI, such as CTransPath \cite{wang2021transpath}, HIPT \cite{chen2022scaling}, and recent models like UNI \cite{chen2024uni}, GigaPath \cite{xu2024whole}, CHIEF \cite{wang2024pathology}, Virchow \cite{vorontsov2024foundation, zimmermann2024virchow2} and H-optimus \cite{hoptimus1}. These foundation models have significantly improved WSI analysis performance. Additionally, task-specific fine-tuning \cite{liu2023multiple,li2023task,tang2024feature} techniques have been introduced to further optimize image feature learning for downstream tasks.

\subsection{Multi-Modal Pretraining in Histopathology}

With the growing availability of multi-modal data and advancements in pretraining techniques across multiple modalities, CLIP-based \cite{radford2021learning,goel2022cyclip,li2023scaling,zhang2024rankclip} vision-language pretraining methods have been applied to text-image pathology datasets, resulting in the development of approaches like MI-zero \cite{lu2023visual}, CATE \cite{huang2024free}, PLIP \cite{huang2023visual}, and Conch \cite{lu2024avisionlanguage}.

Beyond text--image pretraining, gene--WSI integration has also been actively explored using bulk RNA-seq, where a single transcriptomic profile is paired with an entire WSI for slide-level supervision. Representative methods include Pathomic Fusion~\cite{chen2020pathomic}, MCAT~\cite{chen2021multimodal}, SurvPath~\cite{jaume2024modeling}, TANGLE~\cite{jaume2024transcriptomics}, and mSTAR~\cite{xu2025multimodal}. While effective, slide-level supervision is too coarse to align local image regions with their molecular signals. Spatial transcriptomics (ST) overcomes this limitation by providing spot-level paired data, enabling patch-level cross-modal supervision.

Built on ST data \cite{jaume2024hest, chen2024stimage}, recent works span a variety of tasks, such as gene expression prediction~\cite{jaume2024hest, chung2024accurate, chen2024stimage, shi2025multi, huangscalable, zhu2025asign, majumder2026pearl} (e.g., MERGE~\cite{ganguly2025merge}, STEM~\cite{zhudiffusion}, M2OST~\cite{wang2025m2ost}, and Nicheformer~\cite{tejada2025nicheformer}), which maps histology patches to gene expression profiles, and gene--image alignment (e.g., BLEEP~\cite{xie2024spatially}, mclSTExp~\cite{min2024multimodal}, OmiCLIP~\cite{chen2025visual}, and NH\textsuperscript{2}2ST~\cite{qu2025spatially}), which uses gene expression as a supervisory signal to learn better image representations. MoST-IG~\cite{yu2025most} works in the opposite direction, using morphological knowledge to guide gene representation learning. However, existing gene--image alignment approaches treat spots largely independently and may not fully capture cross-spot relationships, nor adequately address the noise and sparsity in gene expression data. In this work, we develop a robust gene--image alignment approach via cross-modal ranking consistency that integrates image and gene features at both local and global scales.

\section{Method}
Our learning pipeline is illustrated in Figure \ref{fig:overview}. The training data consists of pairs of image patches and gene expression profiles, with each pair corresponding to a specific tissue spot. Image patches are fed into image encoders that include both a teacher and a student encoder to facilitate knowledge distillation for gene-guided image feature learning. This setup helps stabilize the image feature learning process during alignment with gene features. Meanwhile, gene expression data are processed by a gene encoder. Following BLEEP~\cite{xie2024spatially} and TANGLE~\cite{jaume2024transcriptomics}, we use the resulting learned gene features as the cross-modal supervisory signal that guides image representation learning.

Both image and gene features are mapped into a shared latent feature space. To align these features, we employ a contrastive learning loss that ensures the image and gene features of the same tissue spot are closely matched. A main contribution is that we introduce a cross-modal ranking consistency loss, ensuring that for any given spot, its similarity to other spots is consistent in both image and gene feature spaces. This ranking loss complements the contrastive loss by providing a global structural constraint over all spot pairs within each mini-batch across the two modalities.

At inference time, we will only use the teacher image encoder to extract gene-guided image features from histopathology image patches. The feature can be used for different downstream tasks that make predictions based on histopathology images. We will first introduce the contrastive learning loss we use for spot-wise alignment, gene-image contrastive (Section \ref{sec:infonce}). Next, we introduce the cross-modal ranking consistency loss (Section \ref{sec:ranking}). We conclude this section with details on the intra-modal distillation for gene-guided image features (Section \ref{sec:distillation}).

\myparagraph{Preliminaries.} Given a set of $N$ spots, $\{1,\ldots,N\}$, each with a pair of image patch $p_n$ (of size $224 \times 224$) and gene profile $\mathbf{x}_n$, we send them into two separate encoders (image encoder $f_\theta$ and gene encoder $h_\phi$), mapping them into image features $i_n = f_\theta(p_n)$ and gene features $g_n = h_\phi(\mathbf{x}_n)$, both of dimension $d = 1024$. $I = \{i_1, i_2, i_3, \dots, i_N\}$ denotes the image features of the $N$ spots, and $G = \{g_1, g_2, g_3, \dots, g_N\}$ their corresponding gene features.

\subsection{Gene-Image Contrastive Loss}
\label{sec:infonce}

To align image and gene features, we first employ the InfoNCE Loss~\cite{oord2018representation}, a commonly used loss function for aligning two modalities. InfoNCE encourages the model to pull positive pairs (image and gene features from the same spot) closer in the shared latent space while pushing apart negative pairs (image and gene features from different spots). 

The Gene-Image Contrastive loss is defined as follows:

\begin{equation}
\mathcal{L}_{\text{gene-image}} = -\sum_{p=1}^N \left[\log \frac{\exp(\text{sim}(i_p, g_p) / \tau)}{\sum_{q=1}^N \exp(\text{sim}(i_p, g_q) / \tau)}\right]
\label{eq:infonce}
\end{equation}
where $\text{sim}(i_\ast, g_\ast)$ is the cosine similarity between the image feature $i_\ast$ and the gene feature $g_\ast$. $\tau$ is a temperature parameter. This loss penalizes unmatched pairs by reducing their similarity while increasing the similarity between matched gene-image pairs. We adopt this one-directional contrastive loss with image features as the anchor to guide image representation learning using gene expression features as the supervisory signal.

\begin{table*}[t]
  \vspace*{-.18in}
  \caption{Gene expression prediction results: (a) top 250 genes and (b) top 250 cancer-specific genes.}
  \label{tab:gene_prediction}
  \centering
  \resizebox{\textwidth}{!}{
  \begin{tabular}{@{}cccccccccc@{}}
    \toprule
    \multirow{2}{*}{\textbf{Image Encoder}} & \multicolumn{3}{c}{\textbf{Breast-ST1}} & \multicolumn{3}{c}{\textbf{Breast-ST2}} & \multicolumn{3}{c@{}}{\textbf{Lung-ST}} \\
    \cmidrule(lr){2-4} \cmidrule(lr){5-7} \cmidrule(lr){8-10}
     & \textbf{MAE} $\downarrow$ & \textbf{MSE} $\downarrow$ & \textbf{PCC} $\uparrow$ & \textbf{MAE} $\downarrow$ & \textbf{MSE} $\downarrow$ & \textbf{PCC} $\uparrow$ & \textbf{MAE} $\downarrow$ & \textbf{MSE} $\downarrow$ & \textbf{PCC} $\uparrow$ \\
    \midrule
    \multicolumn{10}{c}{(a) Results for gene expression prediction (top 250 genes among all genes)} \\
    \midrule
    ResNet-50 \cite{he2016deep} & 0.494 $\pm$ 0.004 & 0.401 $\pm$ 0.006 & 0.069 $\pm$ 0.004 & 0.465 $\pm$ 0.003 & 0.384 $\pm$ 0.008 & 0.069 $\pm$ 0.008 & 0.709 $\pm$ 0.013 & 0.695 $\pm$ 0.019 & 0.026 $\pm$ 0.009 \\
    CTransPath \cite{wang2021transpath} & 0.489 $\pm$ 0.005 & 0.388 $\pm$ 0.012 & 0.088 $\pm$ 0.007 & 0.459 $\pm$ 0.012 & 0.352 $\pm$ 0.013 & 0.132 $\pm$ 0.009 & 0.690 $\pm$ 0.012 & 0.635 $\pm$ 0.017 & 0.039 $\pm$ 0.005 \\
    UNI \cite{chen2024uni} & 0.488 $\pm$ 0.008 & 0.383 $\pm$ 0.010 & 0.096 $\pm$ 0.008 & 0.461 $\pm$ 0.015 & 0.335 $\pm$ 0.017 & 0.154 $\pm$ 0.014 & 0.681 $\pm$ 0.011 & 0.619 $\pm$ 0.019 & 0.043 $\pm$ 0.007 \\
    ST-Net \cite{he2020integrating} & 0.491 $\pm$ 0.004 & 0.395 $\pm$ 0.013 & 0.085 $\pm$ 0.009 & 0.464 $\pm$ 0.012 & 0.362 $\pm$ 0.009 & 0.104 $\pm$ 0.011 & 0.685 $\pm$ 0.015 & 0.615 $\pm$ 0.014 & 0.048 $\pm$ 0.006 \\
    HisToGene \cite{pang2021leveraging} & 0.485 $\pm$ 0.006 & 0.385 $\pm$ 0.017 & 0.102 $\pm$ 0.011 & 0.458 $\pm$ 0.015 & 0.329 $\pm$ 0.015 & 0.144 $\pm$ 0.007 & 0.674 $\pm$ 0.016 & 0.603 $\pm$ 0.015 & 0.055 $\pm$ 0.004 \\
    HEST-FT \cite{jaume2024hest} & 0.482 $\pm$ 0.011 & 0.379 $\pm$ 0.015 & 0.125 $\pm$ 0.013 & 0.452 $\pm$ 0.013 & 0.324 $\pm$ 0.016 & 0.207 $\pm$ 0.005 & 0.661 $\pm$ 0.011 & 0.594 $\pm$ 0.018 & 0.078 $\pm$ 0.004 \\
    BLEEP \cite{xie2024spatially} & 0.477 $\pm$ 0.015 & 0.378 $\pm$ 0.017 & 0.138 $\pm$ 0.009 & 0.454 $\pm$ 0.016 & 0.326 $\pm$ 0.011 & 0.212 $\pm$ 0.006 & 0.669 $\pm$ 0.015 & 0.589 $\pm$ 0.012 & 0.081 $\pm$ 0.001 \\
    NH\textsuperscript{2}2ST \cite{qu2025spatially} & 0.479 $\pm$ 0.011 & 0.381 $\pm$ 0.016 & 0.146 $\pm$ 0.008 & 0.452 $\pm$ 0.014 & 0.325 $\pm$ 0.017 & 0.213 $\pm$ 0.010 & 0.668 $\pm$ 0.019 & 0.591 $\pm$ 0.011 & 0.082 $\pm$ 0.007 \\
    OmiCLIP \cite{chen2025visual} & 0.476 $\pm$ 0.013 & 0.375 $\pm$ 0.012 & 0.154 $\pm$ 0.017 & 0.455 $\pm$ 0.018 & 0.321 $\pm$ 0.015 & 0.218 $\pm$ 0.009 & 0.659 $\pm$ 0.017 & 0.586 $\pm$ 0.011 & 0.083 $\pm$ 0.009 \\
    \emph{RankByGene} & \textbf{0.472} $\pm$ \textbf{0.010} & \textbf{0.371} $\pm$ \textbf{0.012} & \textbf{0.185} $\pm$ \textbf{0.005} & \textbf{0.447} $\pm$ \textbf{0.008} & \textbf{0.318} $\pm$ \textbf{0.010} & \textbf{0.231} $\pm$ \textbf{0.012} & \textbf{0.648} $\pm$ \textbf{0.013} & \textbf{0.570} $\pm$ \textbf{0.011} & \textbf{0.097} $\pm$ \textbf{0.006} \\
    \midrule
    \multicolumn{10}{c}{(b) Results for gene expression prediction (top 250 genes in cancer-specific gene list)} \\
    \midrule
    ResNet-50 \cite{he2016deep} & 0.381 $\pm$ 0.005 & 0.291 $\pm$ 0.008 & 0.089 $\pm$ 0.008 & 0.432 $\pm$ 0.006 & 0.336 $\pm$ 0.005 & 0.112 $\pm$ 0.011 & 0.317 $\pm$ 0.006 & 0.215 $\pm$ 0.007 & 0.041 $\pm$ 0.012 \\
    CTransPath \cite{wang2021transpath} & 0.375 $\pm$ 0.007 & 0.287 $\pm$ 0.005 & 0.098 $\pm$ 0.012 & 0.429 $\pm$ 0.004 & 0.329 $\pm$ 0.007 & 0.132 $\pm$ 0.015 & 0.312 $\pm$ 0.006 & 0.203 $\pm$ 0.009 & 0.047 $\pm$ 0.014 \\
    UNI \cite{chen2024uni} & 0.369 $\pm$ 0.004 & 0.282 $\pm$ 0.009 & 0.116 $\pm$ 0.014 & 0.438 $\pm$ 0.007 & 0.331 $\pm$ 0.005 & 0.127 $\pm$ 0.014 & 0.288 $\pm$ 0.004 & 0.183 $\pm$ 0.005 & 0.054 $\pm$ 0.009 \\
    ST-Net \cite{he2020integrating} & 0.372 $\pm$ 0.007 & 0.284 $\pm$ 0.006 & 0.108 $\pm$ 0.011 & 0.426 $\pm$ 0.006 & 0.327 $\pm$ 0.012 & 0.125 $\pm$ 0.012 & 0.309 $\pm$ 0.007 & 0.198 $\pm$ 0.011 & 0.057 $\pm$ 0.010 \\
    HisToGene \cite{pang2021leveraging} & 0.366 $\pm$ 0.009 & 0.279 $\pm$ 0.011 & 0.134 $\pm$ 0.014 & 0.422 $\pm$ 0.005 & 0.318 $\pm$ 0.016 & 0.146 $\pm$ 0.015 & 0.306 $\pm$ 0.012 & 0.194 $\pm$ 0.010 & 0.064 $\pm$ 0.011 \\
    HEST-FT \cite{jaume2024hest} & 0.362 $\pm$ 0.010 & 0.277 $\pm$ 0.011 & 0.142 $\pm$ 0.011 & 0.417 $\pm$ 0.006 & 0.313 $\pm$ 0.013 & 0.157 $\pm$ 0.019 & 0.301 $\pm$ 0.008 & 0.188 $\pm$ 0.007 & 0.077 $\pm$ 0.007 \\
    BLEEP \cite{xie2024spatially} & 0.363 $\pm$ 0.008 & 0.276 $\pm$ 0.014 & 0.138 $\pm$ 0.013 & 0.415 $\pm$ 0.007 & 0.311 $\pm$ 0.015 & 0.160 $\pm$ 0.009 & 0.297 $\pm$ 0.009 & 0.187 $\pm$ 0.009 & 0.084 $\pm$ 0.005 \\
    NH\textsuperscript{2}2ST \cite{qu2025spatially} & 0.362 $\pm$ 0.021 & 0.278 $\pm$ 0.015 & 0.156 $\pm$ 0.012 & 0.419 $\pm$ 0.011 & 0.314 $\pm$ 0.016 & 0.161 $\pm$ 0.015 & 0.299 $\pm$ 0.011 & 0.188 $\pm$ 0.012 & 0.085 $\pm$ 0.005 \\
    OmiCLIP \cite{chen2025visual} & 0.360 $\pm$ 0.011 & 0.276 $\pm$ 0.015 & 0.169 $\pm$ 0.006 & 0.413 $\pm$ 0.013 & 0.310 $\pm$ 0.017 & 0.166 $\pm$ 0.012 & 0.293 $\pm$ 0.014 & 0.184 $\pm$ 0.005 & 0.089 $\pm$ 0.005 \\
    \emph{RankByGene} & \textbf{0.354} $\pm$ \textbf{0.007} & \textbf{0.272} $\pm$ \textbf{0.009} & \textbf{0.187} $\pm$ \textbf{0.010} & \textbf{0.407} $\pm$ \textbf{0.006} & \textbf{0.306} $\pm$ \textbf{0.012} & \textbf{0.172} $\pm$ \textbf{0.008} & \textbf{0.276} $\pm$ \textbf{0.008} & \textbf{0.176} $\pm$ \textbf{0.005} & \textbf{0.103} $\pm$ \textbf{0.004} \\
    \bottomrule
  \end{tabular}}
   \vspace{-.05in}
   \vspace{-.1in}
\end{table*}

\subsection{Cross-Modal Ranking Consistency Loss}
\label{sec:ranking}

The InfoNCE loss ensures local alignment between image and gene features from the same tissue spot, but it does not address global alignment, which is essential for achieving more accurate and consistent cross-modal correspondences. Directly aligning distances between features from distant tissue spots is not practical, as long-range feature relationships may not be reliable. Instead, we propose that the relative ranking of distances between features is more robust and can provide a more trustworthy basis for alignment.

To leverage this idea, we introduce the cross-modal ranking consistency loss. This loss function encourages the model to learn image representations while maintaining the relative similarity ordering of gene features across tissue spots. By focusing on the ranking of distances rather than exact alignments, the ranking loss facilitates a more reliable and robust global alignment. It complements the local alignment achieved by InfoNCE, while also capturing long-range interactions between features from different tissue spots. In doing so, the ranking consistency loss promotes a more comprehensive and stable cross-modal alignment, improving both local and global correspondences.

Our ranking loss is inspired by the classic ordinal ranking loss \cite{burges2005learning}, but is modified to better address challenges in our gene-image alignment task.
As illustrated in Figure~\ref{fig:ranking}, for a spot of interest, $p$, we want to ensure the similarity between $p$ and two other spots, $q$ and $r$, are ranked consistently in both image feature and gene feature. For simplicity, we denote by $S^I_{p,q}=\text{sim}(i_p,i_q)$ and $S^G_{p,q}=\text{sim}(g_p,g_q)$ the similarity between $p$ and $q$'s image features and gene features. We also denote by $S^I_{p,r}$ and $S^G_{p,r}$ the image and gene feature similarity between $p$ and $r$. Assume $S^G_{p,q} > S^G_{p,r}$. In other words, $g_q$ is closer to $g_p$ compared with $g_r$. We would like to ensure the image features $i_q$ and $i_r$ also maintain the same order. Classic ordinal ranking loss \cite{burges2005learning} will just require the image feature similarity difference $S^I_{p,q} - S^I_{p,r}$ to be bigger than a fixed positive value $\epsilon$. 
In practice, however, we often notice that this loss does not work effectively; image features are often more tightly packed compared with gene features, thus just maintaining the ranking of image similarities will not make many changes, at least in the early alignment stages.

To further accelerate the alignment, we propose a modified ranking loss that pushes image features further apart if possible. In our new ranking loss, we force the image-side similarity gap to be larger than the gene-side gap; this pushes image features apart while preserving the ranking, preventing the image features from collapsing to a uniform space while still respecting the gene-defined ordering.
Formally, we require the difference $S^I_{p,q} - S^I_{p,r}$ to be bigger than $S^G_{p,q} - S^G_{p,r}$. We define 
\begin{equation}
\ell(p, q, r) = \left(S^G_{p, q} - S^G_{p, r}\right) - \left(S^I_{p, q} - S^I_{p, r}\right)
\label{eq:similarity_difference}
\end{equation}
and intend to enforce $\ell(p,q,r)$ to be non-positive for all triplets $(p,q,r)$.
Figure \ref{fig:ranking} illustrates different cases.
In practice, we rewrite the function with the sign function
\begin{equation}
\begin{aligned}
\ell(p, q, r) = \text{sign} & (S^G_{p, q} - S^G_{p, r}) \cdot \\
& \left[\left(S^G_{p, q} - S^G_{p, r}\right) \right. \left. - \left(S^I_{p, q} - S^I_{p, r}\right)\right]
\end{aligned}
\label{eq:sign_similarity_difference}
\end{equation}
This way, even if $S^G_{p,q} < S^G_{p,r}$, requiring $\ell(\cdot)$ to be non-positive will still enforce the relative ranking consistency between image and gene features of the three spots.

Finally, we need a loss to enforce $\ell(\cdot)$ to be nonnegative. To this end, we use a variant of the classic hinge loss, $\max\left\{0,\ell(p,q,r)\right\}$, which incurs penalty if and only if $\ell(p,q,r)$ is positive. Enumerating through all triplets, we have the cross-modal ranking consistency loss

\begin{equation}
\mathcal{L}_{\text{rank}} = \sum_{p}\sum_{q\neq p}\sum_{r\neq p,q} \max\left\{0, \ell(p,q,r)\right\}
\label{eq:ranklosse_e}
\end{equation}

\myparagraph{Accelerating the Computation.}
During training, at each iteration, we will apply the loss to all spots within a mini-batch (batch size $N$). However, this can still be computationally expensive, as it compares $O(N^3)$ pairs of spots. To alleviate the burden, we still enumerate through all $p$'s. But for each $p$, instead of enumerating through all $O(N^2)$ $(q,r)$ pairs, we only sample $O(N)$ random $(q,r)$ pairs. In particular, taking the list of spots except $p$, $L=(1,\ldots,p-1,p+1,\ldots,N)$. We randomly shuffle the sequence, getting a shuffled list, $L'$. We go through the list, $L'$, each time taking two consecutive spots as $q$ and $r$. For the last spot, we will pair it with the first spot in $L'$. This gives us $N-1$ sample $(q,r)$ pairs, and ensures each spot appears in two of the sampled pairs. This reasonably covers similarity rankings of all the spots, but only using $O(N^2)$ triplets.

\subsection{Intra-Modal Distillation Loss}
\label{sec:distillation}
In our framework, we employ a teacher-student network architecture to achieve robust feature representations across differently augmented instances of the same pathology image, drawing on recent advances in self-supervised knowledge distillation for single-modality representation learning \cite{caron2021emerging, oquab2023dinov2}. This intra-modal distillation serves as a stability anchor that complements the cross-modal ranking loss in Section~\ref{sec:ranking}. To enhance stability and invariance in feature embeddings produced by the patch encoder, we apply both weak and strong augmentations to simulate the typical variability found in pathology images. Specifically, the weak view consists of random flipping and cropping, while the strong view further includes color jittering, gaussian blurring, and random grayscale conversion, following established pathology-specific augmentation methods~\cite{kang2023benchmarking}.

In this setup, the weakly augmented image is processed through the teacher encoder, while the strongly augmented version passes through the student encoder. The weights of the teacher encoder are incrementally updated using an Exponential Moving Average (EMA) of the student encoder's weights, which helps stabilize the training. This strategy ensures that the student gradually learns stable features over time.

To enforce the consistency between the representations of the two augmented versions, we introduce the Image Consistency Loss:
\begin{equation}
\mathcal{L}_{\text{distil}} = - \frac{1}{N} \sum_{p=1}^N \left[\log \frac{\exp(\text{sim}(i_{p}^w, i_{p}^s) / \tau)}{\sum_{q=1}^N \exp(\text{sim}(i_{p}^w, i_{q}^s) / \tau)} \right]
\end{equation}
where $N$ is the batch size, $i_{p}^w$ is the feature obtained from the weakly augmented image using the teacher image encoder, and $i_{p}^s$ is the feature from the strongly augmented image using the student image encoder. Minimizing this loss encourages the image encoder to learn representations resilient to augmentation-induced view variations, serving as an intra-modal stability regularizer for the image branch.

\subsection{Overall Loss Function}

The total loss is a weighted sum of the gene-image contrastive loss, the cross-modal ranking consistency loss, and the intra-modal distillation loss:

\begin{equation}
\mathcal{L}_{\text{total}} = \mathcal{L}_{\text{gene-image}} + \lambda_1 \mathcal{L}_{\text{rank}} + \lambda_2 \mathcal{L}_{\text{distil}}
\label{eq:totalloss}
\end{equation}
where \( \lambda_1 \) and \( \lambda_2 \) are hyperparameters that control the balance between the ranking consistency loss and the distillation loss, respectively.

\section{Experiments}
\label{sec:experiments}

We evaluate our method on gene expression prediction, and downstream tasks, including gene mutation prediction, receptor status classification, and survival analysis.

\myparagraph{Datasets and Preprocessing.} To comprehensively evaluate our model's performance, we use the breast and lung ST data from the HEST-1k dataset \cite{jaume2024hest}, which collects publicly available, high-quality ST datasets and applies standardized processing to all collected data. The Breast ST training set consists of 36 samples \cite{andersson2021spatial}, with approximately 15,000 genes per sample; for external evaluation, we use two Visium breast ST samples from 10x Genomics: \renewcommand{\thefootnote}{\arabic{footnote}}{Breast-ST1\footnote{https://www.10xgenomics.com/datasets/human-breast-cancer-visium-fresh-frozen-whole-transcriptome-1-standard} (4,898 spots) and \renewcommand{\thefootnote}{\arabic{footnote}}{Breast-ST2\footnote{https://www.10xgenomics.com/datasets/human-breast-cancer-block-a-section-1-1-standard-1-1-0}} (3,813 spots). The Lung ST training set includes 6 samples \cite{mirzazadeh2023spatially}, each with about 18,000 genes, and the test set comprises 4 lung ST samples (1,831 spots) \cite{villacampa2021genome}. For WSIs, we utilize datasets from \renewcommand{\thefootnote}{\arabic{footnote}}{TCGA\footnote{https://www.cancer.gov/tcga} cohorts} and BCNB \cite{xu2021predicting} for our downstream classification and survival prediction tasks. The pretraining and downstream cohorts cover the same cancer types (breast for BCNB and TCGA-BRCA; lung for TCGA-LUAD), ensuring biological relevance between the pretraining supervisory signal and downstream prediction targets. Following HEST-1k~\cite{jaume2024hest}, we apply a three-step preprocessing pipeline: L1 normalization, log transformation, and 8-neighborhood smoothing to all ST data. The processed data is used in both training and testing.

\begin{table*}[ht]
  \caption{Classification and survival results on BCNB and TCGA Datasets, with slide-level prediction by ABMIL \cite{ilse2018attention}.}
    \subfloat[
    Classification results (AUC) on BCNB and TCGA-LUAD mutation datasets.
    ]{
  \begin{minipage}{0.705\textwidth}
  \scriptsize
    \centering
    \scriptsize
    \setlength{\tabcolsep}{2pt}
  \begin{tabular}{@{}ccccccccc@{}}
    \toprule
    \multirow{2}{*}{\textbf{Image Encoder}} & \multicolumn{3}{c}{\textbf{BCNB}} & \multicolumn{4}{c@{}}{\textbf{TCGA-LUAD mutation}} \\
    \cmidrule(lr){2-4} \cmidrule(lr){5-8}
     & \textbf{BR} & \textbf{PR} & \textbf{HER2} & \textbf{EGFR} & \textbf{KRAS} & \textbf{STK11} & \textbf{TP53} \\
    \midrule
    ResNet-50 \cite{he2016deep} & 0.833 ± 0.039 & 0.785 ± 0.032 & 0.709 ± 0.031 & 0.729 ± 0.068 & 0.643 ± 0.048 & 0.739 ± 0.015 & 0.691 ± 0.023 \\
    CTransPath \cite{wang2021transpath} & 0.857 ± 0.023 & 0.792 ± 0.015 & 0.716 ± 0.019 & 0.812 ± 0.024 & 0.682 ± 0.059 & 0.797 ± 0.025 & 0.783 ± 0.022 \\
    UNI \cite{chen2024uni} & 0.891 ± 0.027 & 0.809 ± 0.018 & 0.724 ± 0.032 & 0.845 ± 0.047 & 0.712 ± 0.046 & 0.843 ± 0.021 & 0.827 ± 0.019 \\
    HEST-FT \cite{jaume2024hest} & 0.901 ± 0.021 & 0.811 ± 0.013 & 0.721 ± 0.015 & 0.851 ± 0.036 & 0.716 ± 0.051 & 0.851 ± 0.028 & 0.834 ± 0.026 \\
    BLEEP \cite{xie2024spatially} & 0.899 ± 0.042 & 0.817 ± 0.019 & 0.734 ± 0.024 & \textbf{0.862 ± 0.028} & 0.713 ± 0.049 & 0.855 ± 0.019 & 0.838 ± 0.029 \\
    \emph{RankByGene} & \textbf{0.915 ± 0.024} & \textbf{0.829 ± 0.012} & \textbf{0.759 ± 0.028} & 0.855 ± 0.037 & \textbf{0.729 ± 0.065} & \textbf{0.861 ± 0.024} & \textbf{0.845 ± 0.033} \\
    \bottomrule
  \end{tabular}
  \label{tab:classification}
  \end{minipage}
  }
  \hfill
    \subfloat[
    Survival analysis results.
    ]{
 \begin{minipage}{0.275\textwidth}
  \scriptsize
  \centering
  \scriptsize
  \setlength{\tabcolsep}{2pt}
  \begin{tabular}{@{}cccc@{}}
    \toprule
    \multirow{2}{*}{\textbf{Image Encoder}} & \multicolumn{2}{c@{}}{\textbf{C-Index}} \\
    \cmidrule(lr){2-3}
     & \textbf{TCGA-BRCA} & \textbf{TCGA-LUAD} \\
    \midrule
    ResNet-50 \cite{he2016deep} & 0.586 $\pm$ 0.052 & 0.563 $\pm$ 0.044 \\
    CTransPath \cite{wang2021transpath} & 0.641 $\pm$ 0.047 & 0.574 $\pm$ 0.036 \\
    UNI \cite{chen2024uni} & 0.668 $\pm$ 0.043 & 0.587 $\pm$ 0.025 \\
    HEST-FT \cite{jaume2024hest} & 0.653 $\pm$ 0.045 & 0.573 $\pm$ 0.032 \\
    BLEEP \cite{xie2024spatially} & 0.672 $\pm$ 0.039 & 0.578 $\pm$ 0.038 \\
    \emph{RankByGene} & \textbf{0.681 $\pm$ 0.051} & \textbf{0.595 $\pm$ 0.029} \\
    \bottomrule
  \end{tabular}
  \label{tab:survival}
  \end{minipage}
  }
  \label{tab:main_result}
  \vspace{-.1in}
  \vspace{-.1in}
\end{table*}

\begin{table*}[ht]
  \caption{Ablation studies of hyperparameters and loss components.}
    \subfloat[Ablation study of $\lambda_1$.
    ]{
  \begin{minipage}{0.195\textwidth}
  \scriptsize
    \centering
    \scriptsize
    \setlength{\tabcolsep}{4pt}
    \begin{tabular}{@{}cccc@{}}
        \toprule
        \multirow{2}{*}{\textbf{$\lambda_1$}} & \multicolumn{3}{c@{}}{\textbf{Breast-ST1}} \\
        \cmidrule(lr){2-4}
         & \textbf{MAE} $\downarrow$ & \textbf{MSE} $\downarrow$ & \textbf{PCC} $\uparrow$ \\
        \midrule
        0  &  0.3603 & 0.2751 & 0.1568\\
        2  &  0.3563 & 0.2735 & 0.1842  \\
        5  &  \textbf{0.3542} & \textbf{0.2723} & \textbf{0.1874}  \\
        10 &  0.3579 & 0.2744 & 0.1819 \\
        \bottomrule
    \end{tabular}
  \label{tab:ablation_study_rank}
  \end{minipage}}
  \hspace{.15in}
    \subfloat[
    Ablation study of $\lambda_2$.
    ]{
  \begin{minipage}{0.195\textwidth}
  \scriptsize
    \centering
    \scriptsize
    \setlength{\tabcolsep}{4pt}
    \begin{tabular}{@{}cccc@{}}
        \toprule
        \multirow{2}{*}{\textbf{$\lambda_2$}} & \multicolumn{3}{c@{}}{\textbf{Breast-ST1}} \\
        \cmidrule(lr){2-4}
         & \textbf{MAE} $\downarrow$ & \textbf{MSE} $\downarrow$ & \textbf{PCC} $\uparrow$ \\
        \midrule
        0   & 0.3611 & 0.2759 & 0.1493  \\
        0.5 & 0.3584 & 0.2747 & 0.1602 \\
        1   & \textbf{0.3542} & \textbf{0.2723} & \textbf{0.1874}  \\
        2   & 0.3592 & 0.2749 & 0.1587 \\
        \bottomrule
    \end{tabular}
  \label{tab:ablation_study_image}
  \end{minipage}}
  \hspace{.15in}
      \subfloat[\centering Ablation study of loss components.
    ]{
 \begin{minipage}{0.26\textwidth}
  \scriptsize
  \centering
  \scriptsize
  \setlength{\tabcolsep}{4pt}
  \begin{tabular}{@{}ccccc@{}}
        \toprule
        \multirow{2}{*}{\textbf{\(\mathcal{L}_{\text{rank}}\)}} & \multirow{2}{*}{\textbf{\(\mathcal{L}_{\text{distil}}\)}} & \multicolumn{3}{c@{}}{\textbf{Breast-ST1}} \\
        \cmidrule(lr){3-5}
         & & \textbf{MAE} $\downarrow$ & \textbf{MSE} $\downarrow$ & \textbf{PCC} $\uparrow$ \\
        \midrule
        \ding{55} & \ding{55} & 0.3624 & 0.2773 & 0.1419 \\
        \checkmark & \ding{55} & 0.3611 & 0.2759 & 0.1493 \\
        \ding{55} & \checkmark & 0.3603 & 0.2751 & 0.1568 \\
        \checkmark & \checkmark & \textbf{0.3542} & \textbf{0.2723} & \textbf{0.1874} \\
        \bottomrule
  \end{tabular}
  \label{tab:loss_component_ablation_study}
  \end{minipage}}
  \hspace{.15in}
      \subfloat[
    Ablation of batch size.
    ]{
 \begin{minipage}{0.25\textwidth}
  \scriptsize
		\centering
  \scriptsize
  \setlength{\tabcolsep}{4pt}
      \begin{tabular}{@{}cccc@{}}
          \toprule
          \multirow{2}{*}{\textbf{Batch Size}} & \multicolumn{3}{c@{}}{\textbf{Breast-ST1}} \\
          \cmidrule(lr){2-4}
          & \textbf{MAE} $\downarrow$ & \textbf{MSE} $\downarrow$ & \textbf{PCC} $\uparrow$ \\
          \midrule
          32  & 0.3591 & 0.2745 & 0.1782 \\
          64  & \textbf{0.3542} & \textbf{0.2723} & \textbf{0.1874} \\
          128 & 0.3582 & 0.2738 & 0.1793 \\
          256 & 0.3596 & 0.2752 & 0.1776 \\
          \bottomrule
      \end{tabular}
  \label{tab:batch_size_ablation}
  \end{minipage}}
  \vspace{-.1in}
  \vspace{-.1in}
  \label{tab:ablation_study_1}
\end{table*}

\begin{table*}[ht]
  \caption{Ablation studies of generalization and robustness.}
  \centering
  \subfloat[
      Ablation of backbone.
      \label{tab:encoder_ablation}
    ]{
      \begin{minipage}{0.275\textwidth}
        \centering
        \scriptsize
        \setlength{\tabcolsep}{4pt}
        \begin{tabular}{@{}cccc@{}}
          \toprule
          \textbf{Encoder} & \textbf{MAE} $\downarrow$ & \textbf{MSE} $\downarrow$ & \textbf{PCC} $\uparrow$ \\
          \midrule
          OmiCLIP~\cite{chen2025visual}     & 0.3604 & 0.2758 & 0.1693 \\
          OmiCLIP + Ours                    & \textbf{0.3421} & \textbf{0.2683} & \textbf{0.1954} \\
          \midrule
          UNI~\cite{chen2024uni}            & 0.3692 & 0.2821 & 0.1163 \\
          UNI + Ours                        & \textbf{0.3542} & \textbf{0.2723} & \textbf{0.1874} \\
          \bottomrule
        \end{tabular}
      \end{minipage}
    }
    \hspace{.05in}
  \subfloat[
    Ablation of initialization.
    \label{tab:initializations}
    ]{
      \begin{minipage}{0.37\textwidth}
        \centering
        \scriptsize
        \setlength{\tabcolsep}{3pt}
        \begin{tabular}{@{}c ccc ccc@{}}
          \toprule
          \multirow{2}{*}{\textbf{Method}} & \multicolumn{3}{c}{\textbf{Virchow2 Init}} & \multicolumn{3}{c}{\textbf{H-optimus-1 Init}} \\
          \cmidrule(lr){2-4} \cmidrule(lr){5-7}
          & MAE $\downarrow$ & MSE $\downarrow$ & PCC $\uparrow$ & MAE $\downarrow$ & MSE $\downarrow$ & PCC $\uparrow$ \\
          \midrule
          HEST-FT & 0.3591  & 0.2775  & 0.1461  & 0.3542 & 0.2733 & 0.1798  \\
          BLEEP   & 0.3563  & 0.2744  & 0.1529  & 0.3515 & 0.2715 & 0.1811  \\
          \emph{RankByGene} & \textbf{0.3476} & \textbf{0.2692} & \textbf{0.1923} & \textbf{0.3426} & \textbf{0.2638} & \textbf{0.2084} \\
          \bottomrule
        \end{tabular}
      \end{minipage}
    }
    \hspace{.05in}
  \subfloat[
    Ablation of sampling strategy.
    \label{tab:sampling}
  ]{
    \begin{minipage}{0.285\textwidth}
      \centering
      \scriptsize
      \setlength{\tabcolsep}{3pt}
      \begin{tabular}{@{}c ccc@{}}
        \toprule
        \multirow{2}{*}{\textbf{Strategy}} & \multicolumn{3}{c}{\textbf{Breast-ST1}} \\
        \cmidrule(lr){2-4}
        & MAE $\downarrow$ & MSE $\downarrow$ & PCC $\uparrow$ \\
        \midrule
        Enumeration ($O(N^3)$) & \textbf{0.3514} & \textbf{0.2694} & \textbf{0.1892} \\ [0.75mm]
        Sampling ($O(N^2)$)    & 0.3542 & 0.2723 & 0.1874 \\ [0.75mm]
        \bottomrule
      \end{tabular}
    \end{minipage}
  }
  \vspace{-.1in}
  \vspace{-.12in}
  \label{tab:ablation_study_2}
\end{table*}

\myparagraph{Baselines.}
We compare our method with image-only pretrained baselines, including ResNet-50 \cite{he2016deep} (pretrained on ImageNet), CTransPath \cite{wang2021transpath}, and UNI \cite{chen2024uni} (all pretrained on pathology images), as well as gene-guided image feature learning baselines HEST-FT \cite{jaume2024hest}, BLEEP \cite{xie2024spatially}, NH\textsuperscript{2}2ST \cite{qu2025spatially}, and OmiCLIP \cite{chen2025visual}. Since HEST-FT, BLEEP, and NH\textsuperscript{2}2ST are gene-image pretraining methods built on configurable image encoders, for a fair comparison, we used UNI for their image encoder initialization. OmiCLIP is itself a pretrained image-omics foundation model and is evaluated using its native encoder. We also compare our method with methods that are specifically designed for gene expression prediction, such as ST-Net \cite{he2020integrating} and HisToGene \cite{pang2021leveraging}, using their default initialization.

\myparagraph{Evaluation Metrics.}
For the gene expression prediction task, we adopt Mean Squared Error (MSE), Mean Absolute Error (MAE), and Pearson Correlation Coefficient (PCC) as evaluation metrics to measure prediction accuracy and correlation, following previous studies \cite{xie2024spatially, chung2024accurate, jaume2024hest}. The WSI classification performance is assessed using the Area Under the Receiver Operating Characteristic Curve (AUC). For survival prediction, we follow \cite{chen2021multimodal, xu2023multimodal, zhou2023cross} and evaluate the model using the Concordance Index (C-Index).

\myparagraph{Implementation Details.}
In the training stage, we use UNI \cite{chen2024uni} as the backbone for the image encoder, followed by a 3-layer MLP as the projection head to map features into a shared space. A 3-layer MLP is also employed for the gene encoder, the same as the previous method \cite{xie2024spatially}. Following \cite{xie2024spatially}, we use cosine similarity to calculate the distance between two modalities. The learning rates are set to $0.0001$ for the image encoder and $0.001$ for the gene encoder. Ranking accuracy is calculated at each epoch using gene embeddings as a convergence criterion. The total number of training epochs is 100, and the batch size is 64. We use the widely-adopted Adam optimizer with its default parameters, EMA momentum of 0.96, and the temperature for gene-image contrastive loss is set to 0.1. All experiments are conducted on an NVIDIA Quadro RTX 8000 GPU (48 GB memory), and each epoch takes about 15 minutes. 

In the testing stage, we perform 5-fold standalone cross-validation for the gene prediction task. The training set is split into five folds, and the best model from each fold is evaluated on the test set. For classification and survival prediction tasks, we conduct standard 5-fold cross-validation and report both the mean and standard deviation. During inference, we use the teacher model as it performs better. 

\myparagraph{Cancer-specific Gene Selection.} Since ST gene expression is high dimensional and extremely sparse, existing methods typically select highly expressed genes for learning and prediction, e.g., top 250 \cite{chung2024accurate}, top 100 \cite{chen2024stimage}, or top 50 \cite{xie2024spatially,jaume2024hest,chung2024accurate} genes. While these highly expressed genes ensure better quality gene features, we are concerned that they may not be relevant to the biology of interest, and thus may be suboptimal for our downstream tasks.

Consulting pathologists, we select prognosis-related genes from the \renewcommand{\thefootnote}{\arabic{footnote}}{Human Protein Atlas\footnote{https://www.proteinatlas.org/humanproteome/cancer}} as our cancer-specific gene lists. This resulted in 447 genes for breast cancer and 1916 genes for lung cancer. These gene lists were used for training and evaluation for downstream tasks, including gene mutation prediction, receptor status classification, and survival analysis. Despite the concern that some of the genes may be close to zero, we observe that these cancer-specific genes provide better performance in downstream tasks. We stress that this should be adapted as the community moves forward with ST data analysis.
Please refer to the supporting document for further details of the gene list. When evaluating on gene prediction task, it is important to avoid sparse genes. We evaluated both the top 250 highly expressed genes overall (as in \cite{chung2024accurate}) and the top 250 highly expressed genes within our cancer-specific gene lists.

\begin{table*}[t]
  \centering
  \small
  \caption{Ablation of (a) the alignment objective and (b) the augmentation strategy on Breast-ST1 (5-fold cross-validation on the cancer-specific 250-gene panel). All variants share the same teacher--student architecture and UNI backbone.}
  \label{tab:align_aug_ablation}
  \begin{minipage}{0.52\textwidth}
    \centering
    {(a) Alignment objective}\\[3pt]
    \setlength{\tabcolsep}{5pt}
    \resizebox{\linewidth}{!}{%
    \begin{tabular}{lcccc}
      \toprule
      \textbf{Method} & \textbf{MAE} $\downarrow$ & \textbf{MSE} $\downarrow$ & \textbf{PCC} $\uparrow$ & \textbf{Complexity} \\
      \midrule
      OT-based~\cite{xu2023multimodal} & $0.356 \pm 0.013$ & $0.274 \pm 0.014$ & $0.174 \pm 0.022$ & $\mathcal{O}(N^{2})$ \\
      Triplet Ranking                  & $0.351 \pm 0.014$ & $0.269 \pm 0.011$ & $0.189 \pm 0.017$ & $\mathcal{O}(N^{3})$ \\
      \midrule
      \textit{RankByGene}              & $0.354 \pm 0.007$ & $0.272 \pm 0.009$ & $0.187 \pm 0.010$ & $\mathcal{O}(N^{2})$ \\
      \bottomrule
    \end{tabular}}
  \end{minipage}
  \hfill
  \begin{minipage}{0.46\textwidth}
    \centering
    {(b) Augmentation strategy}\\[3pt]
    \setlength{\tabcolsep}{6pt}
    \renewcommand{\arraystretch}{1.5}
    \resizebox{\linewidth}{!}{%
    \begin{tabular}{lccc}
      \toprule
      \textbf{Method} & \textbf{MAE} $\downarrow$ & \textbf{MSE} $\downarrow$ & \textbf{PCC} $\uparrow$ \\
      \midrule
      w/o Augmentation                    & $0.355 \pm 0.011$ & $0.273 \pm 0.018$ & $0.179 \pm 0.006$ \\
      \textit{RankByGene} (standard aug.) & $0.354 \pm 0.007$ & $0.272 \pm 0.009$ & $0.187 \pm 0.010$ \\
      \textit{RankByGene} (stain aug.)    & $0.354 \pm 0.016$ & $\mathbf{0.271 \pm 0.007}$ & $\mathbf{0.189 \pm 0.014}$ \\
      \bottomrule
    \end{tabular}}
  \end{minipage}
  \vspace{-.15in}
\end{table*}

\subsection{Quantitative Results}
\myparagraph{Gene Expression Prediction.} As shown in Table \ref{tab:gene_prediction}, \emph{RankByGene} outperforms all baseline methods by a considerable margin across the Breast-ST1, Breast-ST2, and Lung-ST datasets. Specifically, \emph{RankByGene} achieves a PCC improvement of $9\%$ to $34\%$ for the top 250 highly expressed genes and $7\%$ to $35\%$ for the top 250 cancer-specific genes compared to the best baseline performance in each dataset. These results indicate that our method achieves a stronger alignment between image and gene expression data, with the learned gene-informed image features effectively capturing the underlying global relationships between spots.

\myparagraph{Gene-related Classification.} Using models trained on Breast ST and Lung ST datasets, we evaluate their performance on BCNB and TCGA-LUAD mutation datasets, respectively, as shown in Table \ref{tab:main_result} (a). Each method's image encoder extracts patch-level features, which are then aggregated using the ABMIL \cite{ilse2018attention} for slide-level prediction. \emph{RankByGene} achieves strong AUC performance on most mutation statuses, demonstrating that enhanced image-gene alignment enables the model to capture gene-related information more effectively. This highlights the advantage of integrating gene information for downstream tasks.

\myparagraph{Survival Analysis.} Following the same setup, we evaluate survival prediction on TCGA-BRCA and TCGA-LUAD WSIs. Table \ref{tab:main_result} (b) shows that \emph{RankByGene} achieves the highest C-Index (0.6814 for TCGA-BRCA and 0.5945 for TCGA-LUAD), further demonstrating its ability to capture survival-related signals and potential for complex downstream tasks.

\subsection{Ablation Study}

\begin{figure*}[ht]
\centering
\includegraphics[width=1.0\textwidth]{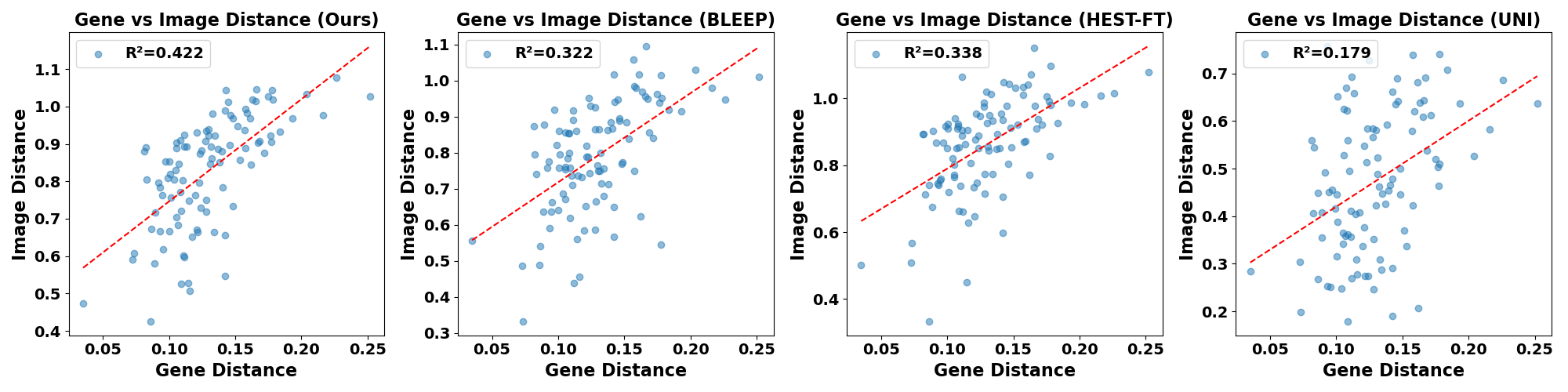}
\vspace{-.2in}
\caption{Comparison of gene--image distances from 100 randomly sampled spot pairs. Each point represents the gene and image distance between two spots. A higher $R^2$~\cite{nagelkerke1991note} indicates stronger linear correlation and better gene--image alignment.}
\label{fig:rankacc2}
\vspace{-.1in}
\end{figure*}

\begin{figure*}[ht]
\centering
\includegraphics[width=0.9\textwidth]{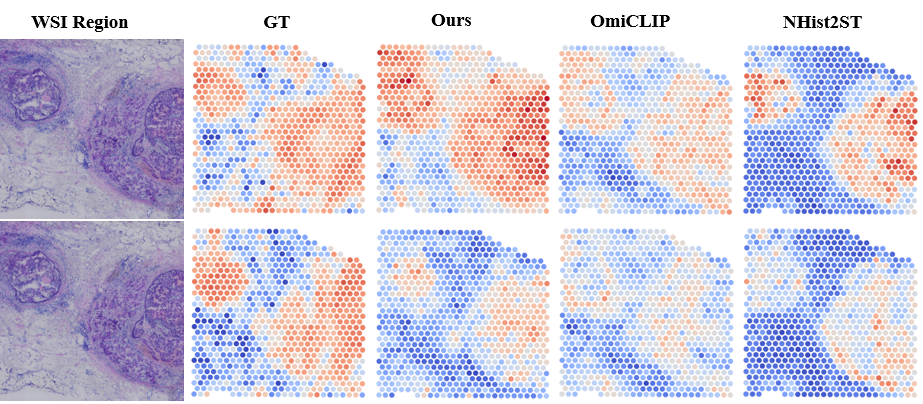}
\caption{Visualization of FASN (first row) and PTGES3 (second row) gene expression predictions from different methods.}
\label{vis}
\vspace{-.1in}
\end{figure*}

\begin{figure}[ht]
\centering
\includegraphics[width=0.47\textwidth]{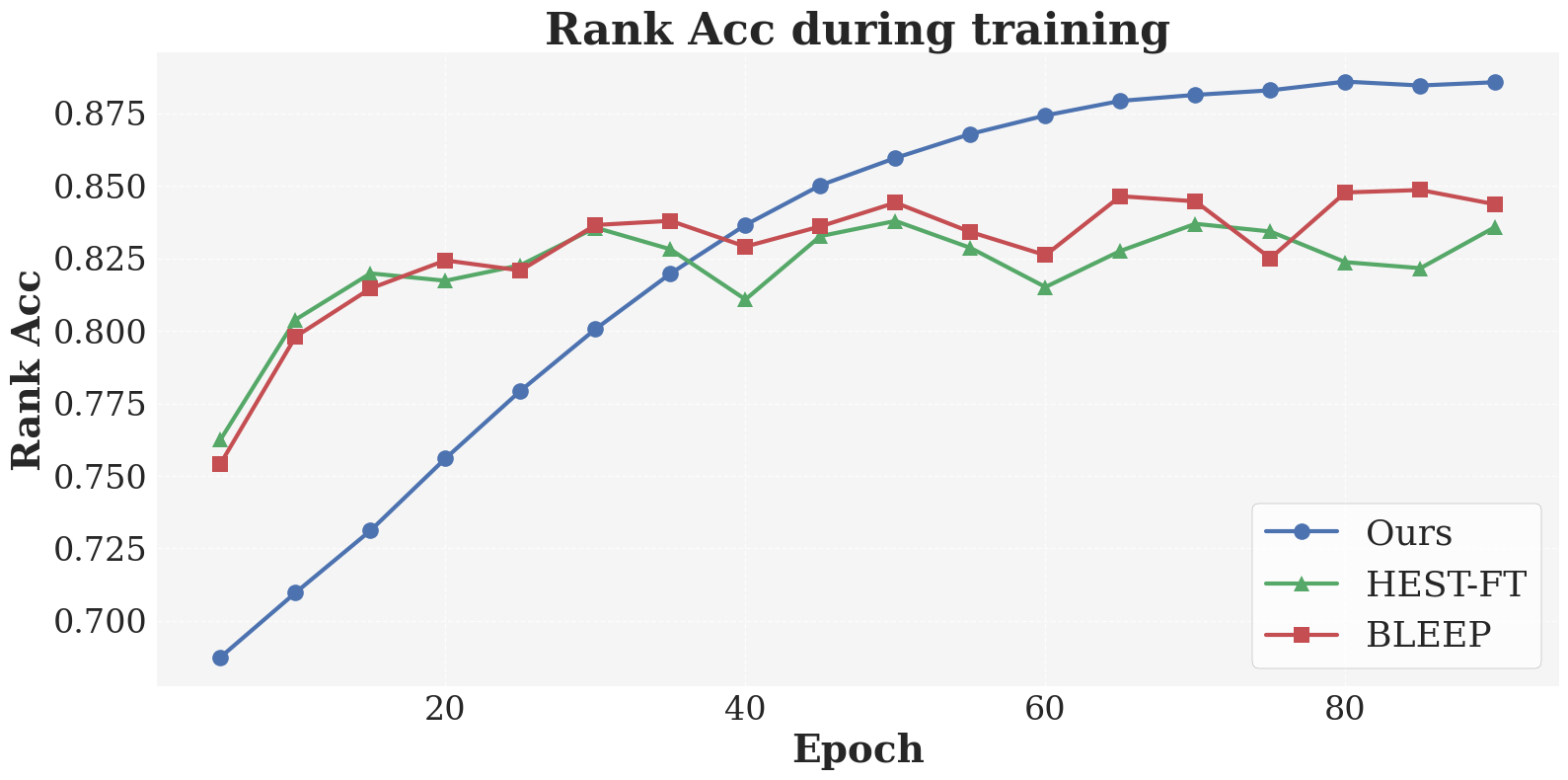}
\vspace{-.1in}
\caption{Rank accuracy for different methods during training.}
\label{fig:rankacc1}
\vspace{-.1in}
\end{figure}

\myparagraph{Hyperparameters and Loss Components.} We conducted ablation studies on key hyperparameters and loss components to evaluate their contributions. We fine-tuned the weights for our two loss terms, $\mathcal{L}_{\text{rank}}$ and $\mathcal{L}_{\text{distil}}$. As shown in Table~\ref{tab:ablation_study_1} (a-b), the optimal balance was achieved with $\lambda_1=5$ and $\lambda_2=1$, as deviating from these values led to performance degradation. Table~\ref{tab:ablation_study_1} (c) confirms the two losses are functionally complementary rather than redundant: each component alone yields only modest gains over the baseline (PCC $0.149$ with $\mathcal{L}_{\text{rank}}$ only, $0.157$ with $\mathcal{L}_{\text{distil}}$ only), while their combination produces a substantially larger gain ($0.187$, a $32\%$ relative improvement over the baseline $0.142$), underscoring the necessity of applying both simultaneously for optimal alignment between gene expression and image features. Our experiments on batch size in Table~\ref{tab:ablation_study_1} (d) identify $64$ as the optimal setting, where the model achieves the highest ranking accuracy at convergence. Performance varies only mildly across the tested range ($\text{PCC}\in[0.178,0.187]$), demonstrating that our method is robust to the batch size choice.

\myparagraph{Generalization and Robustness.}
We further explored our framework's generalization and robustness through ablations on the encoder backbone, initialization schemes, and sampling strategies (Table~\ref{tab:ablation_study_2}). As shown in Table~\ref{tab:ablation_study_2} (a), our approach consistently improves performance across different image encoders, including UNI~\cite{chen2024uni} and OmiCLIP~\cite{chen2025visual}, underscoring its versatility across diverse pretraining strategies. Our method also consistently outperformed baselines across all initialization schemes, including recent pretrained weights such as Virchow2 \cite{zimmermann2024virchow2} and H-optimus-1 \cite{hoptimus1}, demonstrating robustness, stability, and effectiveness regardless of initial parameter settings. Finally, as illustrated in Table~\ref{tab:ablation_study_2} (c), our proposed sampling method for the ranking loss achieves performance comparable to exhaustive enumeration while significantly reducing computational complexity from $O(N^3)$ to $O(N^2)$, thus providing substantial efficiency gains in practice without sacrificing much accuracy.

\myparagraph{Alternative Alignment Objectives and Augmentation.}
We further validate two design choices: the form of the structure-alignment loss and the augmentation pipeline. Table~\ref{tab:align_aug_ablation}~(a) compares our ranking consistency loss against two alternative structure-alignment strategies under the same teacher--student framework and UNI backbone, replacing only the alignment loss: OT-based alignment~\cite{xu2023multimodal} and triplet ranking with exhaustive $\mathcal{O}(N^3)$ enumeration. At the same $\mathcal{O}(N^2)$ complexity as the OT-based variant, \emph{RankByGene} clearly outperforms it (PCC $0.187$ vs.\ $0.174$), and attains accuracy comparable to the much more expensive triplet ranking ($0.187$ vs.\ $0.189$) while reducing per-step complexity by an order of magnitude. Table~\ref{tab:align_aug_ablation}~(b) compares augmentation strategies: a pathology-specific stain augmentation~\cite{kang2023benchmarking} yields only a marginal improvement over our standard pipeline (PCC $0.189$ vs.\ $0.187$), and removing augmentation diversity altogether still retains most of the performance ($0.179$), confirming that the gain is primarily driven by the distillation objective rather than the specific augmentation.

\myparagraph{Encoder Substitution into Existing Frameworks.}
To verify that the representations learned by \emph{RankByGene} are broadly useful beyond our own prediction head, we substitute our encoder into three existing ST prediction frameworks, M2OST~\cite{wang2025m2ost}, MERGE~\cite{ganguly2025merge}, and STEM~\cite{zhudiffusion}, while keeping their prediction architectures unchanged. As shown in Table~\ref{tab:encoder_replacement}, replacing the original encoders with our \emph{RankByGene} encoder consistently improves performance across all three frameworks, demonstrating that our learned features provide stronger inputs for downstream ST prediction regardless of the prediction architecture.

\begin{table}[htbp]
  \centering
  \small
  \setlength{\tabcolsep}{6pt}
  \caption{Encoder substitution experiment on Breast-ST1.}
  \label{tab:encoder_replacement}
  \resizebox{\columnwidth}{!}{%
  \begin{tabular}{lccc}
    \toprule
    \textbf{Method} & \textbf{MAE} $\downarrow$ & \textbf{MSE} $\downarrow$ & \textbf{PCC} $\uparrow$ \\
    \midrule
    M2OST~\cite{wang2025m2ost}         & $0.364 \pm 0.018$ & $0.282 \pm 0.012$ & $0.139 \pm 0.017$ \\
    M2OST + RankByGene         & $\mathbf{0.349 \pm 0.013}$ & $\mathbf{0.271 \pm 0.018}$ & $\mathbf{0.192 \pm 0.015}$ \\
    \midrule
    MERGE~\cite{ganguly2025merge}      & $0.361 \pm 0.015$ & $0.277 \pm 0.016$ & $0.146 \pm 0.012$ \\
    MERGE + RankByGene         & $\mathbf{0.341 \pm 0.017}$ & $\mathbf{0.262 \pm 0.022}$ & $\mathbf{0.212 \pm 0.013}$ \\
    \midrule
    STEM~\cite{zhudiffusion}           & $0.359 \pm 0.014$ & $0.276 \pm 0.013$ & $0.155 \pm 0.016$ \\
    STEM + RankByGene          & $\mathbf{0.346 \pm 0.015}$ & $\mathbf{0.268 \pm 0.010}$ & $\mathbf{0.204 \pm 0.011}$ \\
    \bottomrule
  \end{tabular}}
\end{table}

\begin{figure*}[t]
  \centering
  \includegraphics[width=0.9\linewidth]{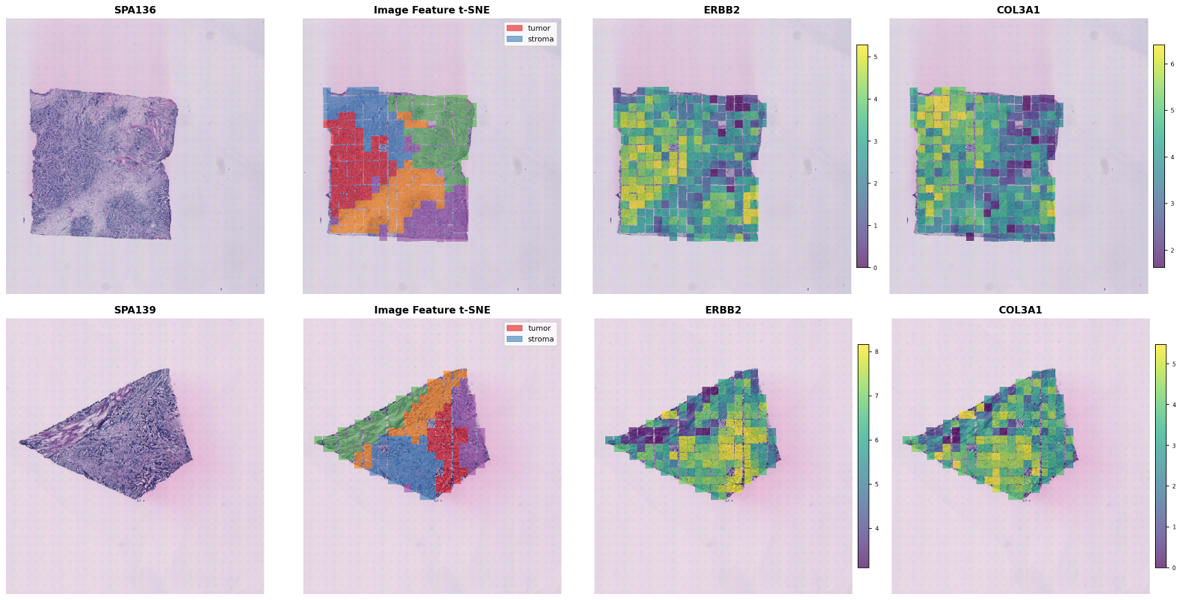}
  \caption{The ranking consistency loss links image features to specific gene programs. Each row shows one representative breast slide (SPA136, SPA139). Columns show the H\&E thumbnail, the t-SNE clustering of image features (\textcolor{red}{red} = ERBB2-enriched cluster, \textcolor{blue}{blue} = COL3A1-enriched cluster), and per-spot expression heatmaps of \textit{ERBB2} and \textit{COL3A1}. The tumor and stromal clusters spatially co-localize with their corresponding marker genes.}
  \label{fig:feature_gene_link}
\end{figure*}

\begin{figure*}[t]
  \centering
  \includegraphics[width=0.9\linewidth]{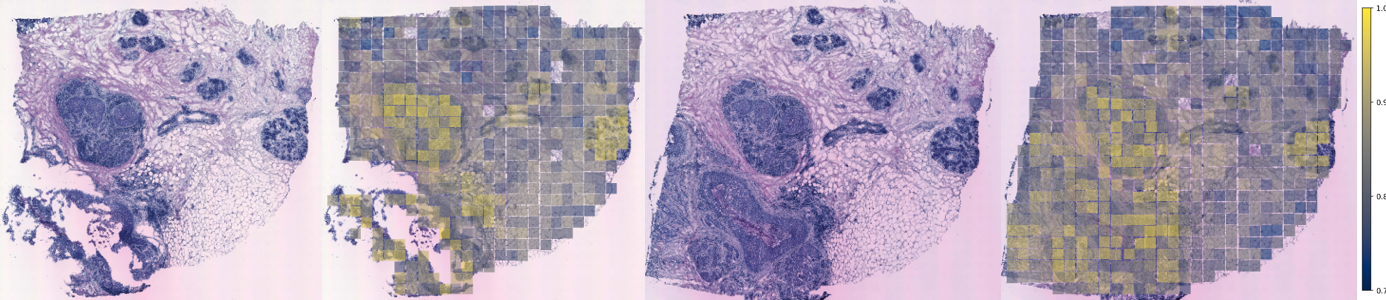}
  \caption{Per-spot gene--image alignment heatmap overlaid on the histology image for two representative slides. Yellow indicates high image--gene alignment, concentrated in tissue-rich regions.}
  \label{fig:alignment_heatmap}
  \vspace{-.1in}
\end{figure*}

\subsection{Qualitative Result}
\myparagraph{Rank Accuracy During Training.}
We define Rank Accuracy as a metric to evaluate whether the ranking relationships among gene features are effectively captured by the image features during training. After each training epoch, Rank Accuracy is calculated as follows: the first gene-image pair in each batch is randomly selected as the anchor pair, along with two additional target pairs. Rank Accuracy is determined by checking whether the target pair with higher gene feature similarity to the anchor pair also has a higher image feature similarity. This process is repeated eight times to calculate the final Rank Accuracy. In Figure \ref{fig:rankacc1}, we compare the change of the rank accuracy during training between different methods. Additionally, in Figure \ref{fig:rankacc2}, we randomly select gene-image pairs to examine whether the distance relationships in gene expression values are reflected in image feature distances. This provides further evidence of the model’s ability to preserve biological relationships in the learned latent representations. Collectively, the two figures demonstrate that our method achieves a significantly larger improvement in accuracy during training compared to the baseline methods, suggesting that the rank relationships in the gene features are more effectively captured by image features.

\myparagraph{Visualization of Cancer Marker Gene.}
To qualitatively evaluate the accuracy of our model’s predictions for clinically relevant genes, we selected FASN and PTGES3 as marker genes for visualization on the Breast-ST1 dataset due to their association with poor prognosis in breast invasive carcinoma across two independent cohorts \cite{he2020integrating}. We visualized the gene prediction values for FASN and PTGES3 from different methods across each spot, including ground truth values and the corresponding WSI region, as shown in Figure \ref{vis}. All prediction values were normalized to the range 0 to 1 to facilitate a direct and fair comparison. Compared to the baseline method, our approach provides visibly more accurate predictions for FASN and PTGES3, demonstrating its effectiveness in capturing biologically meaningful gene expression patterns.

\myparagraph{Linking Image Features to Gene Programs.}
To understand why the ranking mechanism improves cross-modal alignment, we provide a qualitative analysis that directly links the learned image features to specific gene programs. We cluster spots based on image-feature t-SNE coordinates on two representative breast slides (SPA136, SPA139) and recolor each cluster by its dominant marker gene: \textit{ERBB2} (a tumor epithelial marker) and \textit{COL3A1} (a stromal collagen marker). As shown in Figure~\ref{fig:feature_gene_link}, the clusters derived purely from image features spatially align with the expression patterns of these two markers, with the tumor cluster (red) and the stromal cluster (blue) occupying largely non-overlapping regions. This provides a mechanistic explanation for the effectiveness of the ranking consistency loss: by enforcing that the image-side similarity structure mirrors the gene-side similarity structure during training, the loss implicitly pushes the image encoder to organize spots according to their underlying gene programs, so that the image features end up encoding biologically meaningful tissue compartments (tumor vs.\ stroma).

\myparagraph{Gene-Image Alignment Visualization.}
To directly visualize where the learned image features align with gene features at the spot level, we compute a per-spot alignment score (cosine similarity between image and gene features) and overlay it as a heatmap on the corresponding histology image. As shown in Figure~\ref{fig:alignment_heatmap}, high alignment scores (yellow) concentrate in tissue-rich regions, including tumor epithelium and stromal compartments, while low alignment scores (dark) fall on background or sparsely populated areas. This confirms that the image--gene alignment learned by our encoder is strongest in the biologically meaningful regions of the tissue, indicating that the model attends to plausible tissue architecture.

\section{Conclusion}
In this paper, we present a novel, robust, and scalable approach for aligning spatial transcriptomics data with histopathology images, addressing major obstacles in cross-modal alignment. This method employs a ranking-based mechanism to capture similarity relationships between gene and image features, ensuring remarkable resilience to distortions and high effectiveness across both local and global scales while preserving crucial spatial and molecular patterns. A self-supervised knowledge distillation procedure further refines alignment by handling high dimensionality, sparsity, and noise in gene expression data. Through extensive experiments, our framework not only substantially improves alignment quality, but also significantly enhances predictive performance in gene expression prediction, slide-level classification, and survival analysis. This work thus provides a foundational tool and offers valuable new insights for advancing multi-modal representation learning in digital pathology.

\bibliographystyle{ieeetr}
\bibliography{tmi}

\end{document}